\documentclass[preprint,epsfig,graphicx,showpacs,eqsecnum,aps,amsmath,mathrsfs,eufrak,amssymb,floatfix]{revtex4}
\usepackage{graphicx,dcolumn,rotating,booktabs}
\usepackage{epsfig}
\begin{document}

\title{Description of the isotope chain $^{180-196}$Pt within some solvable approaches }

\author{A. A. Raduta$^{a,b}$ and P. Buganu$^{a}$}

\address{$^{a)}$ Department of Theoretical Physics, Institute of Physics and
  Nuclear Engineering,POBox MG6, Bucharest 077125, Romania}

\address{$^{b)}$Academy of Romanian Scientists, 54 Splaiul Independentei, Bucharest 050094, Romania}

\begin{abstract}
Energies of the ground, $\beta$ and $\gamma$ bands as well as the associated B(E2) values are determined for each even-even isotope of the $^{180-196}$Pt chain by the exact solutions of some differential equations which approximate the generalized Bohr-Mottelson Hamiltonian. The emerging approaches are called the Sextic and Spheroidal Approach (SSA), the Sextic and Mathieu Approach (SMA), the Infinite Square Well and Spheroidal Approach (ISWSA) and the Infinite Square Well and Mathieu Approach (ISWMA), respectively. While the first three methods were formulated in some earlier papers of the present authors,  ISWMA is an inedited approach of this work. Numerical results are compared with those obtained with the so called X(5) and Z(5) models. A contour plot for the probability density as function of the intrinsic dynamic deformations is given for a few states of the three considered bands with the aim of evidencing the shape evolution along the isotope chain and pointing out possible shape coexistence.

\end{abstract}

\maketitle

\renewcommand{\theequation}{1.\arabic{equation}}
\setcounter{equation}{0}
\section{Introduction}
\label{sec:level1}
Since the critical point symmetries \cite{Iache2,Iache3,Iache9,Bona1} of the nuclear shape phase transitions were proposed, many experimental and theoretical efforts were made to find the nuclei described by the new symmetries. While at the beginning the X(5) \cite{Iache9} candidates were found in the mass region of $A\approx150$ \cite{CastenZamfir,Krucken,Tonev}, recently a new region has been suggested for Os and Pt isotopes with $A\approx180$ \cite{Dewald,Petkov}. In Refs. \cite{RadBug1,RadBug2} data for the isotopes $^{176,178,180,188,190,192}$Os were analyzed with the Sextic and Spheroidal Approach (SSA) \cite{RadBug1}, the Davidson and Spheroidal Approach (DSA) \cite{RadBug2}, the Infinite Square Well and Spheroidal Approach (ISWSA) \cite{GheRad}  and the results were compared with those of the Coherent State Model (CSM) \cite{RadVal} and X(5). According to our analysis these isotopes present features for  the $U(5)\to SU(3)$ shape phase transition with the critical point reached for $^{176}$Os and $^{188}$Os. On the other hand, applying the Sextic and Mathieu Approach (SMA) \cite{RadBug3} to $^{188,190,192}$Os, one points out that the isotope $^{192}$Os is a good candidate for the critical point of the phase transition between the prolate and the oblate shapes through the triaxial shape corresponding to $\gamma_{0}=30^{0}$.

Encouraged by the results for the Os isotopes, we consider the above mentioned models also for the even-even $^{180-196}$Pt isotopes. We aim not only at determining the energy spectra and the electric transition probabilities but also at showing the static deformation of each isotope in both the ground  and excited states. New features like the shape coexistence or a transition from the prolate to oblate shapes through a triaxial deformation are expected to show up. Keeping in mind that the SMA, the ISWMA and the Z(5) \cite{Bona1} are suitable for the description of 
the triaxial nuclei lying  close to $\gamma_{0}=30^{0}$,  a comparison of their predictions represent a challenging task. ISWMA is the inedited model proposed of this paper.

Recently, in Ref. \cite{Petkov} it was shown that the isotope $^{182}$Pt has some of the X(5) features. According to the Interacting Boson Model-1 (IBM-1) \cite{IacAri} and the General Collective Model \cite{GneGre}, this isotope manifests shape coexistence and it is close to the critical point of the $U(5)\to SU(3)$ shape phase transition. Evidences for shape coexistence were also presented  for $^{176,178}$Pt \cite{Dracoulis1,Dracoulis2}, $^{184}$Pt \cite{Garg}, $^{186}$Pt \cite{Walpe} and $^{188}$Pt \cite{Yuan}, which suggests that this behavior is a specific 
feature for Pt isotopes. Some investigations where the ground state shape evolution in Pt isotope chain from the prolate towards the oblate shapes were performed in Refs. \cite{Nomura,Robledo}.

The objectives formulated above are achieved according to the following plan. In Section II, a short presentation of the formalisms used for the description of the Pt even-even isotopes is given. Numerical results and their comparison with the corresponding experimental data are discussed in Section III. The final conclusions are drawn in Section IV.

\renewcommand{\theequation}{2.\arabic{equation}}
\setcounter{equation}{0}
\section{Short presentation of the models}
\label{sec:level2}
The formalisms X(5), Z(5), ISWSA, ISWMA, SSA and SMA are derived by a set of approximations applied to the Bohr-Mottelson Hamiltonian \cite{Bo},
\begin{equation}
H=-\frac{\hbar^2}{2B}\left[\frac{1}{\beta^4}\frac{\partial }{\partial \beta}
\beta^4 \frac{\partial }{\partial \beta}+\frac{1}{\beta^2\sin {3\gamma} }
\frac{\partial}{\partial \gamma}\sin{3\gamma}\frac{\partial}{\partial \gamma}
-\frac{1}{4\beta^2}\sum_{k=1}^{3}\frac{\hat{Q}_k^2}{\sin^2(\gamma-\frac{2\pi}{3}k)}
\right]+C\frac{\beta^{2}}{2},
\label{Has}
\end{equation}
amended with a potential \cite{Wilets,Fortunato7}
\begin{equation}
V(\beta,\gamma)=V_{1}(\beta)+\frac{V_{2}(\gamma)}{\beta^{2}}.
\end{equation}
The form of the $\beta$ and $\gamma$ potential allows to separate the $\beta$ variable from the  $\gamma$   and the three Euler angles $\theta_{1}$, $\theta_{2}$ and $\theta_{3}$. Here,
$\hat{Q}_{k}$'s  denote the angular momentum components in the intrinsic reference frame. A full separation may be however achieved by expanding the rotor term in power series of $\gamma$  around either of $\gamma_{0}=0$ or of $\gamma_{0}=\pi/6$ and, moreover, by replacing the factor $\beta^2$ multiplying the $\gamma$-dependent term with its average value, denoted hereafter by $\langle\beta^2\rangle$. The resulting equations are:
\begin{equation}
\left[-\frac{1}{\beta^{4}}\frac{d}{d\beta}\beta^{4}\frac{d}{d\beta}+\frac{\Lambda}{\beta^{2}}+v_{1}(\beta)\right]f(\beta)=\varepsilon_{\beta}f(\beta),
\label{eqbeta}
\end{equation}
\begin{equation}
\left[-\frac{1}{\sin3\gamma}\frac{d}{d\gamma}\sin3\gamma\frac{d}{d\gamma}-W+v_{2}(\gamma)\right]\phi(\gamma)=\tilde{\varepsilon}_{\gamma}\phi(\gamma),
\label{eqgamma}
\end{equation}
where the following notations are used:
\begin{equation}
v_{1}(\beta)=\frac{2B}{\hbar^{2}}V_{1}(\beta),\hspace{0.5cm}v_{2}(\gamma)=\frac{2B}{\hbar^{2}}V_{2}(\gamma),\hspace{0.5cm}\varepsilon_{\beta}=\frac{2B}{\hbar^{2}}E_{\beta},\hspace{0.5cm}\tilde{\varepsilon}_{\gamma}=\langle\beta^{2}\rangle\frac{2B}{\hbar^{2}}E_{\gamma}.
\label{enreduse}
\end{equation}
$\Lambda$ and $W$ are the contributions coming from the rotor term and their expressions depend on the order of the $\gamma$ series truncation.

For the sake of fixing the notations and defining the main ingredients, in what follows the above mentioned approaches will be briefly described. For details we advise the reader to consult Refs.
 \cite{Iache9,Bona1,GheRad,RadBug1,RadBug2,RadBug3,Wilets,Fortunato7}.
In Eq. (\ref{eqbeta}), the X(5), Z(5), ISWSA and ISWMA models use a common potential in $\beta$, namely an infinite square well
\begin{equation}
v_{1}(\beta)=\Bigg\{{{0,\,\,\beta\leq\beta_{\omega}} \atop {\infty,\,\,\beta>\beta_{\omega}}}.
\label{udeb}
\end{equation}
With such a choice Eq. (2.3) admits the Bessel functions of irrational order $\nu$, as solutions:
\begin{equation}
f_{s,L}(\beta)=C_{s,L}\beta^{-\frac{3}{2}}J_{\nu}\left(\frac{x_{s,L}}{\beta_{\omega}}\beta\right),\;\;s=1,2,3,....
\end{equation}
$C_{s,L}$ denotes the normalization factor, $x_{s,L}$  the Bessel function zeros, while $L$ is the total intrinsic angular momentum.

By contrast the  SSA and SMA, use in Eq. (\ref{eqbeta}) a sextic oscillator plus a centrifugal barrier potential \cite{Ush},
\begin{equation}
\hspace{0.4cm}v_{1}^{\pm}(\beta)=(b^{2}-4ac^{\pm})\beta^{2}+2ab\beta^{4}+a^{2}\beta^{6}+u_{0}^{\pm},\hspace{0.2cm}c^{\pm}=\frac{L}{2}+\frac{5}{4}+M, \,M=0,1,2,...\,.
\label{v1debeta}
\end{equation}
Here, $c^{\pm}$ has two different values, one for $L$ even and other for $L$ odd, while $u_{0}^{\pm}$ are constants which are fixed such that the two potentials $v_{1}^{+}$ and $v_{1}^{-}$  have the same minimum energy. Eq. (\ref{eqbeta}), with $\Lambda=L(L+1)-2$ and the potential given by Eq. (\ref{v1debeta}), is quasi-exactly solved, the solutions being of the form:
\begin{equation}
\varphi_{n_{\beta},L}^{(M)}(\beta)=N_{n_{\beta},L}P_{n_{\beta},L}^{(M)}(\beta^{2})\beta^{L+1}e^{-\frac{a}{4}\beta^{4}-\frac{b}{2}\beta^{2}},\hspace{0.2cm}n_{\beta}=0,1,2,...M,
\label{sexticbf}
\end{equation}
where $N_{n_{\beta},L}$ is the normalization factor, while $P_{n_{\beta},L}^{(M)}(\beta^{2})$ are polynomials of order $n_{\beta}$ in $\beta^{2}$.

Concerning  Eq.  (\ref{eqgamma}), the X(5) and Z(5) chose an oscillator and a shifted oscillator potential, respectively:
\begin{equation}
v_{2}(\gamma)=c\frac{1}{2}(\gamma-\gamma_{0})^{2}.
\end{equation}
Indeed, for X(5)  $\gamma_0=0$ and the solutions of Eq. (2.4)  are the generalized Laguerre polynomials, $L_{n}^{m}$ :
\begin{equation}
\eta_{n_{\gamma},K}(\gamma)=C_{n,K}\gamma^{|K/2|}e^{-(3a)\gamma^{2}/2}L_{n}^{|K|}(3a\gamma^{2}),
n=\left(\frac{n_{\gamma}-|K|}{2}\right),\,a=\frac{\sqrt{c}}{3},\;\;n_{\gamma}=0,1,2,...
\end{equation}
The quantum number K corresponds to the angular momentum  projection on the intrinsic z-axis.
As for Z(5), $\gamma_{0}=\pi/6$ and the corresponding equation (2.4) is obeyed by
the Hermite polynomials $H_{n}$:
\begin{equation}
\eta_{\bar{n}_{\gamma}}=N_{\bar{n}_{\gamma}}H_{\bar{n}_{\gamma}}(b(\gamma-\pi/6))e^{-b^{2}(\gamma-\pi/6)/2},\;b=\left(\frac{c\langle\beta^{2}\rangle}{2}\right)^{1/4}, \,\bar{n}_{\gamma}=0,1,2,...
\end{equation}
 Both models, the  X(5) and the Z(5), consider in Eq. (\ref{eqgamma}) a zeroth order of approximation for the rotor term. 

This is not the case for the ISWSA, ISWMA, SSA and SMA models, where a second order  power expansion  of both the rotor term  and the periodic potential
\begin{equation}
v_{2}(\gamma)=u_{1}\cos3\gamma+u_2\cos^{2}3\gamma,
\end{equation}
is used, which results of having the spheroidal ($S_{m,n}$) and Mathieu (${\cal M}_{n}$) functions as solutions of the resulting differential equations, respectively:
\begin{equation}
\eta(\gamma)=S_{m,n}(\cos3\gamma;c),\;
\eta(\gamma)=\frac{{\cal M}_{n}(3\gamma;q)}{\sqrt{|\sin3\gamma|}}.
\end{equation}
The expressions of $c$ and $q$ will be specified below.

The advantages of the Mathieu and spheroidal functions consist of that they are periodic, defined on a bound interval and normalized to unity with the integration measure of 
$|\sin3\gamma| d\gamma$, preserving in this way the hermiticity of the initial Hamiltonian. Note that the other approaches do not satisfy these conditions.

The total energy of the system is obtained by summing the eigenvalues of the $\beta$ and $\gamma$ equations:
\begin{equation}
\varepsilon=\varepsilon_{\beta}+\tilde{\varepsilon}_{\gamma}.
\end{equation}
The excitation energies yielded by the formalisms used in the present paper, are as follows:
\begin{equation}
\mbox{{\bf X(5):}}\;E(s,L,n_{\gamma},K)-E(1,0,0,0)=B_{1}(x_{s,L}^{2}-x_{1,0}^{2})+\delta_{K,2}X,\;X=A_1+4C_1,
\end{equation}
with $A_1$, $B_{1}$ and $C_1$ arbitrary parameters. In our calcultions the parameter $X$ is fitted.
\begin{equation}
\mbox{{\bf Z(5):}}\,E(s,L,n_{\gamma}=0,R)-E(1,0,0,0)=B_{1}(x_{s,L,R}^{2}-x_{1,0,0}^{2}),\;B_{1}=\frac{1}{\beta_{\omega}^{2}}\frac{\hbar^{2}}{2B},
\end{equation}
\begin{eqnarray}
\mbox{{\bf ISWSA:}}&&E(s,n_{\gamma},m_{\gamma},L,K)=B_{1}x_{s,L}^{2}+F\left[9\lambda_{m_{\gamma},n_{\gamma}}(c)+\frac{u_{1}}{2}+\frac{11}{27}D-\frac{L(L+1)}{3}\right],\nonumber\\
&&\lambda_{m_{\gamma},n_{\gamma}}=\frac{1}{9}\left[\tilde{\varepsilon}_{\gamma}-\frac{u_{1}}{2}-\frac{11}{27}D+\frac{1}{3}L(L+1)\right],\,c^{2}=\frac{1}{9}\left(\frac{u_{1}}{2}+u_{2}-\frac{2}{27}D\right),\nonumber\\
&&m_{\gamma}=\frac{K}{2},\;\;D=L(L+1)-K^{2}-2,\,F=\frac{1}{\langle\beta^{2}\rangle}\frac{\hbar^{2}}{2B},
\end{eqnarray}
\begin{eqnarray}
\mbox{{\bf ISWMA:}}&&E(s,n_{\gamma},L,R)=B_{1}x_{s,L}^{2}+F\left[9a_{n_{\gamma}}(L,R)+18q(L,R)-\frac{3}{4}R^{2}-\frac{5}{2}\right],\\
&&q=\frac{1}{36}\left(\frac{10}{9}L(L+1)-\frac{13}{12}R^{2}+u_{1}-\frac{9}{4}\right),\,a_{n_{\gamma}}=\frac{1}{9}\left(\tilde{\varepsilon}_{\gamma}+\frac{3}{4}R^{2}+\frac{5}{2}\right)-2q,\nonumber
\end{eqnarray}
\begin{eqnarray}
&&\mbox{{\bf SSA:}}\nonumber\\
&&E(n_{\beta},n_{\gamma},m_{\gamma},L,K)=G\left[b(2L+3)+\lambda_{n_{\beta}}^{(M)}+u_{0}^{\pm}\right]+F\left[9\lambda_{m_{\gamma},n_{\gamma}}(c)+\frac{u_{1}}{2}+\frac{11}{27}D-L(L+1)\right],\nonumber\\
&&\lambda_{m_{\gamma},n_{\gamma}}=\frac{1}{9}\left[\tilde{\varepsilon}_{\gamma}-\frac{u_{1}}{2}-\frac{11}{27}D+\frac{1}{3}L(L+1)\right]+\frac{2L(L+1)}{27},\,G=\frac{\hbar^{2}}{2B},
\end{eqnarray}
\begin{eqnarray}
&&\mbox{{\bf SMA:}}\\
&&E(n_{\beta},n_{\gamma},L,R)=G\left[4bs(L)+\lambda_{n_{\beta}}^{(M)}(L)+u_{0}^{\pi}\right]+F\left[9a_{n_{\gamma}}(L,R)+18q(L,R)-\frac{3}{4}R^{2}-\frac{5}{2}\right]\nonumber,
\end{eqnarray}
where $\lambda_{n_{\beta}}^{(M)}(L)$ satisfy the equation:
\begin{equation}
\left[-\left(\frac{d^{2}}{d\beta^{2}}+\frac{4s-1}{\beta}\frac{d}{d \beta}\right)+2b\beta\frac{d}{d\beta}+2a\beta^{2}\left(\beta\frac{d}{d \beta}-2M\right)\right]P_{n_{\beta},L}^{(M)}(\beta^{2})=\lambda_{n_{\beta}}^{(M)}P_{n_{\beta},L}^{(M)}(\beta^{2}).
\label{Qoper}
\end{equation}
The specific $\beta$ and $\gamma$ potentials of the six approaches used in the present paper are collected, for comparison, in Table \ref{tab-0}. The potentials in the $\beta$ variable are to be amended by a centrifugal term due to the rotor component of the starting Hamiltonian.
\begin{table}
\caption{ Here we list the $\beta$ and $\gamma$ potentials used by the approaches having as acronyms ISWSA, SSA, ISWMA and SMA. For comparison the potentials characterizing X(5) and Z(5) are also given.}
\begin{tabular}{|c|l|l|}
\hline
Approach       & $\beta$ potential &  $\gamma$ potential
\\ \hline 
X(5)           & $0, \;\;\rm{for}\;\; \beta\leq\beta_{\omega};$&  $ \frac{c}{2}\gamma^2.$\\
               & $\infty,\;\;\rm{for}\;\;\beta>\beta_{\omega}.$&                         \\  
\hline
Z(5)           &$0, \;\;\rm{for}\;\; \beta\leq\beta_{\omega}$;&   $\frac{c}{2}(\gamma -\frac{\pi}{6})^2.$\\ 
               & $\infty,\;\;\rm{for}\;\;\beta>\beta_{\omega}.$& \\
\hline
ISWSA          &$0, \;\;\rm{for}\;\; \beta\leq\beta_{\omega};$ &    $u_1\cos3\gamma+u_2\cos^23\gamma+\frac{9}{4\sin^23\gamma}.$\\
               & $\infty,\;\;\rm{for}\;\;\beta>\beta_{\omega}.$& \\
\hline
SSA            &$(b^2-4ac^{\pm})\beta^2+2ab\beta^4+a^2\beta^6+u^{\pm}_0,$&  $u_1\cos3\gamma+u_2\cos^23\gamma+\frac{9}{4\sin^23\gamma}.$\\
               &$c^{\pm}=\frac{L}{2}+\frac{5}{4}+m;\;m=0,1,2,...$        &    \\ 
\hline
ISWMA          &$0 \;\;\rm{for}\;\; \beta\leq\beta_{\omega};$& $-2q\cos6\gamma ;$\\
               & $\infty,\;\;\rm{for}\;\;\beta>\beta_{\omega}.$&$ q=\frac{1}{36}\left(\frac{10}{9}L(L+1)-\frac{13}{12}R^2+\mu-\frac{9}{4}\right). $                       \\  
\hline
  SMA          &$(b^2-4ac^{\pm})\beta^2+2ab\beta^4+a^2\beta^6+u^{\pm}_0,$& $-2q\cos6\gamma;$\\
               & $c^{\pm}=\frac{L}{2}+\frac{5}{4}+m;\;m=0,1,2,...$  &  $q=\frac{1}{36}\left(\frac{10}{9}L(L+1)-\frac{13}{12}R^2+\mu-\frac{9}{4}\right).$                                                   \\
\hline
\end{tabular}
\label{tab-0}
\end{table}

The reduced E2 transition probabilities for ISWSA and SSA are determined with the reduced matrix element of the transition operator:
\begin{equation}
T^{(E2)}_{2\mu}=t_1\beta \left[\cos\gamma D^2_{\mu 0}+\frac{\sin\gamma}{\sqrt{2}}(D^2_{\mu 2}+D^2_{\mu, -2})\right]+t_2\sqrt{\frac{2}{7}}\beta^2 \left[-\cos2\gamma D^2_{\mu 0}+\frac{\sin2\gamma}{\sqrt{2}}(D^2_{\mu 2}+D^2_{\mu, -2})\right],
\label{transop}
\end{equation}
between the corresponding initial $|L_iM_i\rangle$ and final $|L_fM_f\rangle$ states, as described above:
\begin{equation}
B(E2;L_{i}\rightarrow L_{f})=|\langle L_{i}||T_{2}^{(E2)}||L_{f}\rangle|^{2}.
\label{redtrans}
\end{equation}
Here the Rose's convention \cite{Rose} was used for the reduced matrix elements.
For SMA, ISWMA and Z(5), in the expression of the transition operator (\ref{transop}) $\gamma$ is substituted with $\gamma-2\pi/3$. The argument is justified by the fact that $\gamma-2\pi/3$ defines the axis 1 of the principal inertial ellipsoid. Indeed, the transformation from the laboratory to the intrinsic frame is a rotation defined by the matrix $D^{L}_{MR}$, where  M and R are eigenvalues of the operator $\hat{Q}_1$. The X(5) and Z(5) models keep only the zero order approximation of the first $\gamma$-term in the transition operator (\ref{transop}).

\renewcommand{\theequation}{3.\arabic{equation}}
\setcounter{equation}{0}
\section{Numerical results}
\label{sec:level3}
The formalisms presented in Section II were applied to some even-even isotopes of Pt: $^{180-196}$Pt. It is commonly accepted that nuclear spectra can be classified by the values of the energy ratios:
\begin{eqnarray}
R_{4_{g}^{+}/2_{g}^{+}}&=&\frac{E_{4_{g}^{+}}-E_{0_{g}^{+}}}{E_{2_{g}^{+}}-E_{0_{g}^{+}}},\nonumber\\
R_{0_{\beta}^{+}/2_{g}^{+}}&=&\frac{E_{0_{\beta}^{+}}-E_{0_{g}^{+}}}{E_{2_{g}^{+}}-E_{0_{g}^{+}}}.
\label{ratios}
\end{eqnarray} 
Moreover, it seems that nuclei satisfying a certain symmetry are characterized by almost constant ratios. The values of these ratios associated to the isotopes considered here are collected in Table II. As seen from there, the ratios $R_{4_{g}^{+}/2_{g}^{+}}$ for $^{180,182,184}$Pt are close to that predicted by the $X(5)$ approach. By contrast the other ratio
$R_{0_{\beta}^{+}/2_{g}^{+}}$ indicates that these isotopes are far from the ideal picture of $X(5)$. As a matter of fact this feature is consistent with the results of Ref.\cite{RadGheFa}
saying that not all nuclear properties reach the critical point in a phase transition in the same isotope. We apply the approaches ISWSA and SSA to the mentioned isotopes in order to test their ability to account for these complementary features.

Concerning the description called $Z(5)$ this is appropriate for $^{190,192,194,196}$Pt, the statement being supported by the values of both ratios. Indeed, the detailed numerical analysis of Ref.\cite{Bona1} shows a good agreement between  calculations and experimental data. In this context the application of the ISWMA and SMA to these isotopes will provide a sensible comparison of the formalisms on one hand and the $Z(5)$ on the other hand.

It is well known that the triaxial  rigid rotor (TRR) predicts \cite{Davydov} a relation between the first three excited state energies:
\begin{equation}
\bigtriangleup E\equiv E_{3_{\gamma}^{+}}-E_{2_{\gamma}^{+}}-E_{2_{g}^{+}}=0.
\end{equation}
Due to this fact the above equation is considered to be a signature for a triaxial deformation of $\gamma_{0}=30^{0}$. For the isotope $^{192}$Pt the above equation reads:
$|\Delta E| =8$ keV, which means that the mentioned isotope is close to the ideal triaxial rigid rotor. Considering this isotope among the treated isotopes allows us to answer the question whether these approaches are suitable for the description of the triaxial nuclei. The isotopes $^{186,188,190,192,194,196}$Pt may be considered to be $\gamma-$unstable nuclei, having the ratio $R_{4_{g}^{+}/2_{g}^{+}}$ close to 2.5. A special case is that of $^{186}$Pt  which has the head state of gamma band higher in energy than the  first beta state which results in claiming a gamma stable picture. Most likely this nucleus exhibits the main features for the critical point of the phase transition  from prolate to oblate shapes. Due to the specific structure of their potentials in the $\gamma$ variable, the 
ISWSA and SSA seem to be suitable to describe both the $\gamma-$unstable and $\gamma-$stable deformed nuclei. Actually this argument justifies including $^{186}$Pt and $^{188}$Pt on the list of considered isotopes. In addition to the prolate-oblate transition along the Pt isotopic chain an alternative prolate-oblate transition has been considered in Ref. \cite{Jolie}, with both transitions studied in \cite{Bona5}.
\begin{table}[h!]
\caption{Signatures of X(5), Z(5) and O(6) symmetries identified in the even-even isotopes $^{180-196}$Pt. The two ratios are defined by Eq.(\ref{ratios}). }
\vspace{0.3cm}
\begin{tabular}{|c|cccccccccccc|}
\hline
                        &$^{180}$Pt&$^{182}$Pt& $^{184}$Pt& $^{186}$Pt &$^{188}$Pt&$^{190}$Pt& $^{192}$Pt&$^{194}$Pt&$^{196}$Pt&X(5)& Z(5) & O(6) \\
\hline
$R_{4_{g}^{+}/2_{g}^{+}}$&2.69 &2.71 &2.67 &2.55 & 2.52 &2.49 & 2.48 & 2.47 &2.46 & 2.90 & 2.35 & 2.50 \\
$R_{0_{\beta}^{+}/2_{g}^{+}}$& 3.12& 3.23& 3.02& 2.46& 3.00& 3.11& 3.78 & 3.86 & 3.19& 5.65& 3.91 &  - \\
\hline
\end{tabular}
\label{tab-1}
\end{table}

Each approach involves a number of free parameters for energies as well as for B(E2) values. These are fixed by fitting some particular experimental data concerning either the excitation energies or the reduced transition probabilities. The results of the fitting procedure adopted are listed in Tables \ref{tab-2}$,  $\ref{tab-3}. As seen from these tables, the number of parameters used for fitting the spectra in $X(5)$, $ISWSA$  and $SSA$ are 2, 4, 6, respectively, while  in the fitting of $B(E2)$ values 1, 2, 2 parameters are used, respectively. Also, from Table III we notice that the number of parameters used for fitting the spectra in Z(5), ISWMA and SMA are 1, 3, 5 respectively, while in fitting the $B(E2)$s 1, 2, 2 parameters are used respectively.
\begin{table}
\caption{The parameters characterizing the X(5), ISWA and SSA approaches, determined by a fitting procedure, are listed for $^{180-188}$Pt isotopes. }
\scriptsize{
\begin{tabular}{|c|ccccccccccccccccc|}
\hline
Nucl& \multicolumn{2}{c}{$B_1$[keV]}&X[keV]&\multicolumn{2}{c}{ F[keV]}&\multicolumn{2}{c}{$u_1$}&\multicolumn{2}{c}{$u_2$}&G[keV]&a& b&\multicolumn{3}{c}{$t_1$[W.u.]$^{1/2}$}&
\multicolumn{2}{c}{$t_2$[W.u.]$^{1/2}$}\\ 
\cline{2-18}
       & X(5) & ISWSA & X(5) & ISWSA & SSA & ISWSA &SSA  &ISWSA   &SSA  &SSA &SSA&SSA & X(5) & ISWSA & SSA &  ISWSA &  SSA \\
\hline
$^{180}$Pt&19.08&16.38&722.5&17.32&3.34  &-0.15& -821.2& -104.6&-1000&1.04 &1059&135&500.2& 614.4 &1750 &0.0 &0.0\\
$^{182}$Pt&18.02&16.39&720.7&11.33&5.33  &-31.56& -1042&  -163 &-0.0007&0.81&1687&186&451.2&2200&6561  &9062 &89567\\
$^{184}$Pt&17.28&16.83&739.7& 3.35&6.25  &-1000 & -302.6& -1000&-262&0.62&3030&256&419.6 &2422&7821  &11331&122065\\
$^{186}$Pt&     &16.25&     &16.82&3.08  &-253.87&1471&   6.75&-2326&0.85&1296&170&      &1728&5061  &5978&58515\\
$^{188}$Pt&     & 21.5&     &41.99&14.55 &-97.45 & -466.2 &81.07 &165.8&1.45&1449&95&    &517.4&1717 &0.0&0.0\\
\hline
\end{tabular}}
\label{tab-2}
\end{table}
\begin{table}
\caption{The parameters characterizing the Z(5), ISWMA and SMA approaches, determined by a fitting procedure, are listed for $^{190-196}$Pt isotopes. }
\scriptsize{
\begin{tabular}{|c|cccccccccccccc|}
\hline
Nucl& \multicolumn{2}{c}{$B_1$[keV]}&\multicolumn{2}{c}{ F[keV]}&\multicolumn{2}{c}{$u_1$}&G[keV]&a& b&\multicolumn{3}{c}{$t_1$[W.u.]$^{1/2}$}&\multicolumn{2}{c}{$t_2$[W.u.]$^{1/2}$}
\\ \cline{2-15}
       & Z(5) & ISWMA & ISWMA & SMA &ISWMA & SMA &ISWMA  &SMA& SMA& Z(5) & ISWMA & SMA & ISWMA &  SMA \\
\hline
$^{190}$Pt&28.12&16.73&12.82&8.14&26.67& 104.6& 1.11&3014.12&84.00&27.49&28.14  &96.38  &0.00  &0.0   \\
$^{192}$Pt&29.45&17.84&13.98&7.87&9.49 & 121.8&2.95 &616.5&22.98& 23.94&24.51 &55.10  &102.4 &1048 \\
$^{194}$Pt&32.65&19.87&18.43&14.68&5.00 & 32.74& 2.96&733.0&33.05&18.76&   16.94 &43.42 &137.6 & 968.6\\
$^{196}$Pt&31.49&18.27&9.98 &6.48&56.53& 177.1& 0.41&28322&250&20.77&     19.79 &130.2  &172.9 &7708  \\
\hline
\end{tabular}}
\label{tab-3}
\end{table}

Numerical results for the excitation energies of the ground, $\beta$ and $\gamma$ bands, as well as  for the quadrupole electric transitions between states of these bands are compared with the corresponding experimental data in Tables \ref{tab-4} and \ref{tab-5}, respectively. For each formalism the agreement between the calculation results and the corresponding experimental data is quantitatively appraised by the r.m.s values of the deviations. 

From Table \ref{tab-4}, one can see that  spectra of the isotopes $^{180}$Pt, $^{182}$Pt and $^{184}$Pt are better described  by SSA and ISWSA  than by X(5). The best approach seems to be SSA. Moreover, the  X(5) failure in explaining the data from the $\beta$ band is removed by SSA, and that happens especially for $^{182}$Pt. Concerning the $\gamma$ band, all three formalisms, SSA, ISWSA and X(5), encounter difficulties in explaining the band head energy. A possible explanation would be that the state $2_{\gamma}^{+}$,  does not actually belong  to the $\gamma$ band. In this context we mention the fact that two alternative interpretations have been studied in Ref. \cite{McC1} with a related description  appearing in Ref.\cite{McC2}. A similar situation is met in $^{186}$Pt. In $^{188}$Pt, however, all three bands considered here are realistically described  by SSA. 

The comparison of the numerical results yielded by SMA, ISWMA and Z(5) with experimental data for the even-even isotopes $^{190-196}$Pt, is made also in Table \ref{tab-4}, with the result in favor of SMA and ISWMA.

\begin{sidewaystable}[htbp!]
\caption{Excitation energies, given in units of keV,  of the ground, $\beta$ and $\gamma$ bands states J$_{i}^{+}$ with $i=g,\beta,\gamma$,   yielded by the SSA, ISWSA, X(5) for $^{180-188}$Pt   and SMA ISWMA and Z(5) approaches for $^{190-196}$Pt, are compared with the corresponding experimental data taken from Refs. 
\cite{WuNiu,BalrajJoel,Baglin1,Baglin2,Balraj1,Balraj2,Baglin3,Balraj3,Huang}. The r.m.s. values of the prediction deviations from the corresponding experimental data, denoted $\chi$ and given in units of keV, are also listed}.
\begin{flushleft}
\resizebox{24cm}{8cm} {
\footnotesize{
\scriptsize{
\scriptsize{
\begin{tabular}{|c|cccc|cccc|cccc|ccc|ccc|cccc|cccc|cccc|cccc|cccc|}
\hline
  &\multicolumn{4}{|c|}{$^{180}$Pt}&\multicolumn{4}{|c|}{$^{182}$Pt}&\multicolumn{4}{|c|}{$^{184}$Pt}&\multicolumn{3}{|c|}{$^{186}$Pt}&\multicolumn{3}{|c|}{$^{188}$Pt}&\multicolumn{4}{|c|}{$^{190}$Pt}&\multicolumn{4}{|c|}{$^{192}$Pt}&\multicolumn{4}{|c|}{$^{194}$Pt}&\multicolumn{4}{|c|}{$^{196}$Pt}\\
\hline
J$_{i}^{+}$&Exp&SSA&ISWSA&X(5)&Exp&SSA&ISWSA&X(5)&Exp&SSA&ISWSA&X(5)&Exp&SSA&ISWSA&Exp&SSA&A&Exp&SMA&ISWMA&Z(5)&Exp&SSA&ISWMA&Z(5)&Exp&SSA&ISWMA&Z(5)&Exp&SMA&ISWMA&Z(5)\\
\hline
$2_{g}^{+}$ & 153  &  126 & 125  & 133  & 155  & 139  & 121  & 126  & 163  & 131  & 119  & 121& 192  & 146  & 123  & 266  & 232  & 183  & 296  & 225  & 282  & 284& 317  & 214  & 303  & 297  & 328  & 252  & 334  & 329  & 356  & 255  & 314  & 318  \\
$4_{g}^{+}$ & 411  &  386 &  366 & 387  & 420  & 412  & 353  & 366  & 436  & 393  & 347  & 351& 490  & 426  & 362  & 671  & 645  & 545  & 737  & 645  & 721  & 667& 785  & 647  & 772  & 698  & 811  & 723  & 835  & 774  & 877  & 748  & 824  & 747  \\
$6_{g}^{+}$ & 757  &  749 &  693 & 724  & 775  & 778  & 666  & 684  & 798  & 749  & 650  & 656& 878  & 801  & 685  & 1185 & 1170 & 1045 & 1288 & 1206 & 1259 & 1130& 1365 & 1247 & 1346 & 1184 & 1412 & 1347 & 1435 & 1313 & 1526 & 1414 & 1466 & 1266   \\
$8_{g}^{+}$ & 1182 & 1194 & 1093 & 1131 & 1206 & 1216 & 1047 & 1069 & 1231 & 1176 & 1018 & 1025& 1343 & 1250 & 1080 & 1783 & 1772 & 1667 & 1915 & 1872 & 1885 & 1668  & 2018 & 1979 & 2010 & 1747 & 2100 & 2081 & 2120 & 1936 & 2253 & 2215 & 2227 & 1868 \\
$10_{g}^{+}$& 1674 & 1705 & 1563 & 1604 & 1698 & 1710 & 1492 & 1515 & 1707 & 1658 & 1445 & 1453& 1858 & 1757 & 1543 & 2438 & 2429 & 2405 & 2628 & 2620 & 2591 & 2276& 2729 & 2820 & 2759 & 2383 & 2848 & 2899 & 2883 & 2642 & 3044 & 3125 & 3101 & 2548 \\
$12_{g}^{+}$& 2229 & 2273 & 2100 & 2139 & 2242 & 2251 & 1999 & 2021 & 2204 & 2185 & 1929 & 1938&2336 & 2315 & 2073 & 3105 & 3127 & 3256 &      &      &      & &&&&&&&&&&&&  \\
$14_{g}^{+}$& 2842 & 2891 & 2702 & 2736 & 2832 & 2830 & 2568 & 2585 & 2727 & 2749 & 2470 & 2478& 2825 & 2916 & 2667 &      &      &      &      &      &      & &&&&&&&&&&&&       \\
$16_{g}^{+}$& 3505 & 3552 & 3369 & 3392 & 3461 & 3442 & 3195 & 3205 & 3282 & 3344 & 3066 & 3073 & 3395 & 3556 & 3325 &      &      &      &      &      &      &&&&&&&&&&&&&   \\
$18_{g}^{+}$& 4253 & 4253 & 4099 & 4108 & 4094 & 4083 & 3882 & 3881 & 3869 & 3967 & 3716 & 3721& 4051 & 4229 & 4045 &      &      &      &      &      &      &&&&&&&&&&&&&    \\
$20_{g}^{+}$& 4985 & 4989 & 4892 & 4882 & 4729 & 4749 & 4627 & 4613 & 4493 & 4611 & 4420 & 4422& 4788 & 4933 & 4827 &      &      &      &      &      &      &&&&&&&&&&&&&    \\
$22_{g}^{+}$& 5729 & 5757 & 5748 & 5715 & 5403 & 5437 & 5430 & 5400 & 5167 & 5276 & 5178 & 5176& 5597 & 5666 & 5671 &      &      &      &      &      &      &&&&&&&&&&&&&      \\
$24_{g}^{+}$& 6551 & 6555 & 6663 & 6605 & 6127 & 6143 & 6290 & 6241 & 5897 & 5957 & 5988 & 5983& 6464 & 6424 & 6575 &      &      &      &      &      &      &&&&&&&&&&&&&  \\
$26_{g}^{+}$& 7434 & 7379 & 7641 & 7552 & 6905 & 6867 & 7208 & 7136 & 6686 & 6652 & 6852 & 6841& 7408 & 7205 & 7540 &      &      &      &      &      &      & &&&&&&&&&&&&  \\
$28_{g}^{+}$&      &      &      &      &      &      &      &      & 7535 & 7360 & 7767 & 7751&      &      &      &      &      &      &      &      &      & &&&&&&&&&&&&\\
\hline
$0_{\beta}^{+}$&  478 & 590  &  649 & 753  & 500  & 537  & 647  & 712  & 492  & 581  & 665  & 682 & 472  & 472  & 642  & 799  & 719  & 849  & 921  & 832  & 661  & 1110& 1195  & 1108  & 705  & 1163 & 1267 & 1150 & 785  & 1289 & 1135 & 948  & 722  & 1244  \\
$2_{\beta}^{+}$&  861 & 809  &  863 & 993  & 856  & 797  & 860  & 939  & 844  & 822  & 878  & 900  & 798  & 743  & 856  & 1115 & 1193 & 1153 & 1203 & 1260 & 1173 & 1617& 1439  & 1489  & 1254 & 1693 & 1512 & 1623 & 1392 & 1877 & 1362 & 1428 & 1288 & 1810  \\
$4_{\beta}^{+}$& 1248 & 1173 & 1258 & 1425 & 1240 & 1185 & 1246 & 1347 & 1234 & 1198 & 1263 & 1291  & 1223 & 1134 & 1247 &      & 1802 & 1716 &      & 1875 & 1931 & 2259&       & 2117  & 2062 & 2366 &      & 2328 & 2272 & 2623 &      & 2138 & 2145 & 2530  \\
$6_{\beta}^{+}$& 1650 & 1632 & 1760 & 1967 & 1650 & 1652 & 1734 & 1859 & 1800 & 1655 & 1747 & 1782 & 1600 & 1604 & 1744 &      & 2493 & 2446 &      & 2607 & 2815 & 2999&       & 2903  & 3005 & 3140 &      & 3165 & 3283 & 3482 &      & 2996 & 3165 & 3358   \\
$8_{\beta}^{+}$&      & 2164 & 2348 & 2593 & 2118 & 2180 & 2303 & 2450 &      & 2173 & 2307 & 2348 &      & 2135 & 2325 &      & 3240 & 3314 &      & 3426 & 3803 & 3822&       & 3807  & 4055 & 4003 &      & 4094 & 4398 & 4438 &      & 3969 & 4322 & 4280 \\
$10_{\beta}^{+}$&     & 2755 & 3013 & 3292 &      & 2755 & 2943 & 3111 &      & 2738 & 2934 & 2982 &      & 2718 & 2982 &      & 4028 & 4308 &      &      & 4885 & 4724 &&&&&&&&&&&&\\
\hline
$2_{\gamma}^{+}$& 677   & 840  & 858  &  856 & 668  & 805  & 849  & 847  & 649  & 817  & 859  & 860 & 607  & 849  & 917  & 606  & 681  & 723  & 598  & 648  & 581  & 521& 612  & 668  & 552  & 546  & 622  & 627  & 638  & 605  & 689  & 724  & 660  & 584  \\
$3_{\gamma}^{+}$& 963   & 954  & 969  &  971 & 943  & 924  & 955  & 956  & 940  & 932  & 962  & 965  & 957  & 970  & 1027 & 936  & 860  & 887  & 917  & 848  & 812  & 737& 921  & 877  & 798  & 772  & 923  & 868  & 909  & 856  & 1015 & 951  & 915  & 825 \\
$4_{\gamma}^{+}$& 1049  & 1101 & 1105 & 1110 & 1034 & 1079 & 1084 & 1087 & 1028 & 1080 & 1087 & 1090 & 992  & 1130 & 1161 & 1085 & 1098 & 1089 & 1128 & 1159 & 1183 & 1254& 1201 & 1184 & 1201 & 1313 & 1229 & 1284 & 1378 & 1456 & 1293 & 1280 & 1290 & 1405\\
$5_{\gamma}^{+}$& 1315  & 1258 & 1263 & 1269 & 1306 & 1236 & 1234 & 1237 & 1307 & 1234 & 1230 & 1235 & 1363 & 1290 & 1317 &      & 1316 & 1325 & 1450 & 1369 & 1391 & 1315& 1482 & 1418 & 1418 & 1377 & 1499 & 1492 & 1590 & 1527 & 1610 & 1543 & 1560 & 1473 \\
$6_{\gamma}^{+}$&       & 1464 & 1440 & 1447 & 1438 & 1446 & 1402 & 1405 & 1463 & 1438 & 1391 & 1396 & 1470 & 1505 & 1492 & 1636 & 1630 & 1594 & 1733 & 1808 & 1882 & 2004& 1869 & 1865 & 1953 & 2099 &      & 2090 & 2214 & 2327 & 2007 & 2008 & 2051 & 2245 \\
$7_{\gamma}^{+}$& 1727  & 1653 & 1637 & 1642 & 1731 & 1630 & 1587 & 1589 & 1731 & 1617 & 1567 & 1572 & 1801 & 1693 & 1687 &      & 1868 & 1893 &      & 2009 & 2062 & 1949& 2113 & 2106 & 2134 & 2041 &      & 2246 & 2360 & 2263 &      & 2286 & 2328 & 2183\\
$8_{\gamma}^{+}$&       & 1909 & 1853 & 1854 &      & 1886 & 1789 & 1790 &      & 1866 & 1759 & 1764 & 2004 & 1954 & 1899 & 2247 & 2241 & 2223 &      & 2559 & 2665 & 2799& 2591 & 2678 & 2792 & 2931 &      & 3004 & 3130 & 3250 & 2750 & 2870 & 2928 & 3134 \\
$9_{\gamma}^{+}$& 2198  & 2122 & 2087 & 2082 &      & 2088 & 2008 & 2005 &      & 2064 & 1965 & 1971& 2280 & 2163 & 2129 &      & 2489 & 2583 &      & 2742 & 2816 & 2644&      & 2914 & 2938 & 2769 &      & 3095 & 3211 & 3070 &      & 3151 & 3211 & 2961  \\
$10_{\gamma}^{+}$&      & 2421 & 2338 & 2326 &      & 2382 & 2243 & 2236 &      & 2351 & 2186 & 2192& 2545 & 2462 & 2377 &      & 2911 & 2971 &      & 3391 & 3222 & 3647 &     & 3597 & 3371 & 3819 &      & 3995 & 3665 & 4234 &      & 3841 & 3694 & 4084  \\
\hline
$\chi$     &      & 58   & 108  & 128  &      & 47   & 156  & 164  &      & 83   & 155  & 151  &      & 107  & 155  &      &  45  & 89   &      & 67   &   97 & 218&      & 76   & 158  & 193  &      & 69   & 160  & 157  &      & 92   & 135  & 274 \\
\hline
\end{tabular}}}}}
\end{flushleft}
\label{tab-4}
\end{sidewaystable}

The electromagnetic transition probabilities, calculated with Eq. (\ref{redtrans}), are included in Tables \ref{tab-5}. Analyzing the r.m.s. values for each model, one may conclude that SSA and ISWSA  describe the experimental data batter than X(5), while SMA and ISWMA better than Z(5). An explanation for this picture could be  that X(5) and Z(5) use only the zero order approximation of the harmonic part of the transition operator (\ref{transop}). Indeed,   as shown in Table \ref{tab-5}, for $^{180}$Pt the results obtained by SSA and ISWSA using only the harmonic transition operator are almost identical with those of X(5). By contrast for $^{182,184}$Pt where the anharmonic contributions were included, the results of SSA and ISWSA are better than those of X(5). It is worth noticing that  the r.m.s. associated to Z(5)  for  $^{192}$Pt and $^{196}$Pt are smaller than those provided by ISWMA. This situation might be caused by the fact the two approached considered for the $\gamma$ band different descriptions. Indeed, in the framework of Z(5) the states of $\gamma$ band are characterized by $n_{\gamma}=0$, while the ISWMA $\gamma$ states have $n_{\gamma}=1$.

\begin{sidewaystable}[htbp!]
\caption{The reduced E2 transition probabilities determined with the SSA, ISWSA and X(5) models for $^{180-188}$Pt and SMA, ISWMA and Z(5) for $^{190-196}$Pt, are compared with the corresponding  experimental data taken from Refs. \cite{WuNiu,Walpe,Baglin1,Balraj1,Balraj2,Baglin3,Balraj3,Huang}. }
\begin{flushleft}
\resizebox{24cm}{8cm}{
\footnotesize{
\scriptsize{
\scriptsize{
\begin{tabular}{|c|cccc|cccc|cccc|ccc|ccc|cccc|cccc|cccc|cccc|}
\hline
B(E2) [W.u.]&\multicolumn{4}{|c|}{$^{180}$Pt}&\multicolumn{4}{|c|}{$^{182}$Pt}&\multicolumn{4}{|c|}{$^{184}$Pt}&\multicolumn{3}{|c|}{$^{186}$Pt}&\multicolumn{3}{|c|}{$^{188}$Pt}&\multicolumn{4}{|c|}{$^{190}$Pt}&\multicolumn{4}{|c|}{$^{192}$Pt}&\multicolumn{4}{|c|}{$^{194}$Pt}&\multicolumn{4}{|c|}{$^{196}$Pt}\\
\hline
J$_{i}^{+}\rightarrow$J$_{f}^{'+}$&Exp&SSA&ISWSA&X(5)&Exp&SSA&ISWSA&X(5)&Exp&SSA&ISWSA&X(5)&Exp&SSA&ISWSA&Exp&SSA&A&Exp&SMA&ISWMA&Z(5)&Exp&SSA&ISWMA&Z(5)&Exp&SSA&ISWMA&Z(5)&Exp&SMA&ISWMA&Z(5)\\
\hline
$2_{g}^{+}\rightarrow0_{g}^{+}$   & $153^{+15}_{-15}$ & 110 & 106 & 106 & $108^{+7}_{-7}$   & 167 & 166 & 86  & $127^{+5}_{-5}$   & 176 & 179 & 75 & $113^{+8}_{-8}$   & 162 & 162 & $82^{+15}_{-15}$ & 82  & 82  & $56^{+3}_{-3}$ & 56  & 56  & 56& $57.2^{+1.2}_{-1.2}$ & 49 & 41  & 42 & $49.2^{+0.8}_{-0.8}$ & 25 & 20 & 26 & $40.6^{+0.2}_{-0.2}$ & 34 & 28 & 32 \\
$4_{g}^{+}\rightarrow2_{g}^{+}$   & $140^{+30}_{-30}$ & 168 & 169 & 169 & $188^{+11}_{-11}$ & 226 & 222 & 138 & $210^{+8}_{-8}$   & 238 & 236 & 119& $188^{+13}_{-13}$ & 232 & 228 &                  & 136 & 131 &                & 86  & 95  & 89& $89^{+5}_{-5}$       & 73 & 71  & 68 & $85^{+5}_{-5}$       & 37 & 34 & 41 & $60^{+0.9}_{-0.9}$   & 52 & 48 & 51 \\
$6_{g}^{+}\rightarrow4_{g}^{+}$   & $\geq50$          & 202 & 210 & 210 & $284^{+18}_{-18}$ & 232 & 224 & 171 & $226^{+12}_{-12}$ & 243 & 235 & 148&$289^{+23}_{-23}$ & 254 & 248 &                  & 171 & 162 &                & 119 & 138 & 123 &  $70^{+30}_{-30}$     & 98 & 103 & 94 & $67^{+21}_{-21}$     & 51 & 49 & 57 & $73^{+4}_{-73}$      & 72 & 70 & 70 \\
$8_{g}^{+}\rightarrow6_{g}^{+}$   &                   & 230 & 241 & 241 & $253^{+20}_{-20}$ & 221 & 215 & 196 & $271^{+18}_{-18}$ & 232 & 222 & 170& $294^{+29}_{-29}$ & 260 & 253 &                  & 200 & 186 &                & 144 & 169 & 148 &                      &    &     &    & $50^{+14}_{-14}$     & 61 & 60 & 69 & $78^{+10}_{-78}$     & 87 & 85 & 84  \\
$10_{g}^{+}\rightarrow8_{g}^{+}$  &                   & 255 & 265 & 266 & $266^{+21}_{-21}$ & 204 & 202 & 216 & $310^{+10}_{-10}$ & 214 & 205 & 187& $304^{+26}_{-26}$ & 259 & 254 &                  & 226 & 205 &                & 166 & 191 & 166&                      &    &     &    & $34^{+9}_{-9}$       & 70 & 68 & 77 &                      &    &    &   \\
$12_{g}^{+}\rightarrow10_{g}^{+}$ &                   & 278 & 285 & 286 & $158^{+18}_{-18}$ & 185 & 189 & 232 & $183^{+17}_{-17}$ & 193 & 188 & 201& $255^{+26}_{-26}$ & 252 & 252 &                  & 249 & 220 &                &     &     &   &&&&&&&&&&&&   \\
$14_{g}^{+}\rightarrow12_{g}^{+}$ &                   & 300 & 301 & 302 & $113^{+11}_{-11}$ & 164 & 178 & 246 & $165^{+17}_{-17}$ & 171 & 173 & 213& $225^{+21}_{-21}$ & 243 & 249 &                  &     &     &                &     &     &   &&&&&&&&&&&&  \\
$16_{g}^{+}\rightarrow14_{g}^{+}$ &                   &     &     &     &                   &     &     &     & $143^{+17}_{-17}$ & 150 & 159 & 223& $201^{+36}_{-36}$ & 232 & 246 &                  &     &     &                &     &     &  &&&&&&&&&&&& \\
$18_{g}^{+}\rightarrow16_{g}^{+}$ &                   &     &     &     &                   &     &     &     & $80^{+5}_{-5}$    & 129 & 147 & 231&&&&&&&&&& &&&&&&&&&&&&\\
\hline
$2_{\beta}^{+}\rightarrow0_{\beta}^{+}$ &&&&&&&&&&  &  &  &  &  &  &  &  & &&&&&&&&&&&& & $5^{+5}_{-5}$ & 23 & 20 & 25 \\
\hline
$3_{\gamma}^{+}\rightarrow2_{\gamma}^{+}$ &&&&&&&&&& &&&&&&&&&&&& & $102^{+10}_{-10}$ & 89 & 85 & 92 &                &    &    &    &                  &    &   &  \\
$4_{\gamma}^{+}\rightarrow2_{\gamma}^{+}$ &&&&&&&&&&&&&&&&&&&&&&  &                   &    &    &    & $21^{+4}_{-4}$ & 15 & 14 & 19 & $29^{+6}_{-29}$  & 22 & 19 & 24 \\
$6_{\gamma}^{+}\rightarrow4_{\gamma}^{+}$&&&&&&&&&& &&&&&&&&&&&& &                   &    &    &    &                &    &    &    & $49^{+13}_{-13}$ & 29 & 28 & 33 \\
\hline
$0_{\beta}^{+}\rightarrow2_{g}^{+}$ &&&&&&&&&&  &&&&&&&&&&&&    &  &  &  &  & $0.63^{+0.14}_{-0.14}$ & 9.13 & 21.43 & 19.55 & $2.8^{+1.5}_{-1.5}$          & 15.5 & 30.8   & 24 \\
$2_{\beta}^{+}\rightarrow0_{g}^{+}$ &&&&&&&&&&  &&&&&&&&&&&&    &  &  &  &  &                        &      &       &       & $0.0025^{+0.0024}_{-0.0024}$ &0.18  & 0.0033 & 0.34 \\
$2_{\beta}^{+}\rightarrow4_{g}^{+}$ &&&&&&&&&&  &&&&&&&&&&&&    &  &  &  &  &                        &      &       &       & $0.13^{+0.12}_{-0.12}$       & 9.8  & 13.6   &  10.5 \\
$0_{\beta}^{+}\rightarrow2_{\gamma}^{+}$&&&&&&&&&& &&&&&&&&&&&& &  &  &  &  & $8.4^{+1.9}_{-1.9}$    & 39.9 & 1.9   & 0     & $18^{+10}_{-10}$             & 21   & 1 & 0 \\
$2_{\beta}^{+}\rightarrow2_{\gamma}^{+}$&&&&&&&&&& &&&&&&&&&&&& &  &  &  &  &                        &      &       &      & $0.26^{+0.23}_{-0.23}$       & 0.02 & 8 & 4.6 \\
\hline
$2_{\gamma}^{+}\rightarrow0_{g}^{+}$    &&&&&&&&&& &&&&&&&&&&&& & $0.55^{+0.04}_{-0.04}$ & 0.93 & 3.42 & 0  & $0.29^{+0.04}_{-0.04}$ & 1.29 & 1.75 & 0  &                       &  &  &  \\
$2_{\gamma}^{+}\rightarrow2_{g}^{+}$    &&&&&&&&&& &&&&&&&&&&&& &                        &      &      &    & $89^{+11}_{-11}$       & 71   &  87  & 42 &                       &  &  &  \\
$3_{\gamma}^{+}\rightarrow2_{g}^{+}$    &&&&&&&&&& &&&&&&&&&&&& & $0.68^{+0.07}_{-0.07}$ & 1.74 & 7.13 & 0  &                        &      &      &    &                       &  &  &  \\
$3_{\gamma}^{+}\rightarrow4_{g}^{+}$     &&&&&&&&&&&&&&&&&&&&&& & $38^{+10}_{-10}$       & 38   & 38   & 53 &                        &      &      &    &                       &  &  &  \\
$4_{\gamma}^{+}\rightarrow2_{g}^{+}$    &&&&&&&&&& &&&&&&&&&&&& &                        &      &      &    & $0.36^{+0.07}_{-0.07}$ & 0.79 & 1.21 & 0  & $0.56^{+0.12}_{-0.17}$& 0.32& 0.69 & 0 \\
$4_{\gamma}^{+}\rightarrow4_{g}^{+}$    &&&&&&&&&& &&&&&&&&&&&& &                        &      &      &    & 14                     & 16   & 21   & 9  &                       &  &  &  \\
$6_{\gamma}^{+}\rightarrow4_{g}^{+}$    &&&&&&&&&& &&&&&&&&&&&& &                        &      &      &    &                        &      &      &    & $0.48^{+0.14}_{-0.14}$& 0.19 & 0.41 & 0 \\
$6_{\gamma}^{+}\rightarrow6_{g}^{+}$    &&&&&&&&&& &&&&&&&&&&&& &                        &      &      &    &                        &      &      &    & $16^{+5}_{-5}$        & 5    & 9    & 6 \\
\hline
r.m.s. [W.u.]                     &                   &  36 & 39  & 39  &                   & 47  & 52  & 80  &                   & 43  & 49  & 86  &                   & 36  & 40  &                  &     &     &                &     &     &  &                        & 14   & 17   & 15 &                        & 22   & 22   & 25 &                       & 9 & 13 & 11  \\
\hline
\end{tabular}}}}}
\end{flushleft}
\label{tab-5}
\end{sidewaystable}
\begin{table}
\caption{The branching ratios for some states of the $^{188}$Pt and $^{190,192,194}$Pt isotopes determined with SSA, ISWSA and SMA, ISWMA, Z(5), respectively, are compared with the corresponding experimental data taken from Ref. \cite{Finger}.}
\vspace{0.3cm}
{\scriptsize
\resizebox{16cm}{3.8cm} {
\begin{tabular}{|c||c|c|c||c|c|c|c||c|c|c|c||c|c|c|c|}
\hline
$\frac{B(E2;J^{+}\rightarrow J^{'+})}{B(E2;I^{+}\rightarrow I^{'+})}$&\multicolumn{3}{|c||}{$^{188}$Pt}&\multicolumn{4}{|c||}{$^{190}$Pt}&\multicolumn{4}{|c|}{$^{192}$Pt}&\multicolumn{4}{|c|}{$^{194}$Pt}\\
\hline
$\times 10^{2}$&Exp.&SMA&ISWSA&Exp.&SMA&ISWMA&Z(5)&Exp.&SMA&ISWMA&Z(5)&Exp.&SMA&ISWMA&Z(5)\\
\hline
$\frac{2_{\gamma}^{+}\rightarrow0_{g}^{+}}{2_{\gamma}^{+}\rightarrow2_{g}^{+}}$     &3.44    &63    &66    &1.24&1.95&4.90 &0   &0.51 &1.96&7.55&0&0.38&1.81&2.01&0\\
$\frac{3_{\gamma}^{+}\rightarrow2_{g}^{+}}{3_{\gamma}^{+}\rightarrow2_{\gamma}^{+}}$&4.5     &23    &17    &1.8 &2.2 &6.0  &0   &0.76     &1.95&8.42&0&0.5 &5.37&9.04&0\\
$\frac{3_{\gamma}^{+}\rightarrow4_{g}^{+}}{3_{\gamma}^{+}\rightarrow2_{\gamma}^{+}}$&        &9.9   &7.3   &49  &49  &49   &57  &26       &43  &45&57&    &128&182&57\\
$\frac{0_{\beta}^{+}\rightarrow2_{g}^{+}}{0_{\beta}^{+}\rightarrow2_{\beta}^{+}}$   &$\geq$11&23.2  &17    &11  &14  &31   &19  &3.8      &8.1 &31&19&7.9 &10.5&31&19\\
$\frac{2_{\beta}^{+}\rightarrow0_{g}^{+}}{2_{\beta}^{+}\rightarrow0_{\beta}^{+}}$   &0.83    &0.37  &3.17  &0.02&0.82&0.02 &1.39&0.022    &1.16&0.017&1.39&    &1.04&0.02&1.39\\
$\frac{2_{\beta}^{+}\rightarrow4_{g}^{+}}{2_{\beta}^{+}\rightarrow0_{\beta}^{+}}$   &19      &66    &49    &4.2 &44  &68   &42  &$\leq$2.8&28  &68&42&    &35&68&42\\
\hline
r.m.s.                                                                              &        &35    &32    &    &16  &27   &16&         &13  &30&22&    &3&14&7\\
\hline
$t_{2}$ [W.u.]$^{\frac{1}{2}}$                                                      &        &-316.8&-184.9&    &2931&145.2&    & &&        &&&&    &\\
\hline
\end{tabular}}}
\label{tab-6}
\end{table}

In Table VII we list the results for branching ratios of few states from the $\gamma$ and $\beta$ bands
obtained by SSA, ISWSA, SMA and ISWMA approaches, respectively. They are compared
with the experimental data of Ref.[40]. For $^{190,192,194}$Pt we list also the results yielded by
the Z(5) formalism. The parameters determining the transition operator were fixed as follows.
For $^{188}$Pt and $^{190}$Pt we kept $t_1$ as given in Tables III and IV respectively, while $t_2$ was fixed
by a least square procedure. The results for $t_2$ are also listed in Table VII. As for the rest
of isotopes from the mentioned Table, the parameters $t_1$, $t_2$ are as listed in Tables III, IV.

Another objective of the present work is to determine the isotope shape in ground and excited states, within both the  SSA and the SMA. Indeed, it is interesting to see how the shape changes when one passes from one isotope to another and moreover whether this picture is state dependent.We expect to visualize the shape phase transition and also possible shape coexistence. The static shape is defined by the values of the intrinsic variables $\beta$ and $\gamma$ for which the probability density (the probability in the volume unit of $d\beta d\gamma$),
\begin{equation}
P(\beta,\gamma)=|f(\beta)\phi(\gamma)|^{2}\beta^{4}|\sin3\gamma|,
\end{equation}
reaches a maximum value. In Fig.1, the contour plots are represented in the coordinates $(\beta\cos\gamma ,\beta\sin\gamma)$. In order to save the space we chose two representatives for SSA, $^{180}$Pt and $^{188}$Pt, and one for SMA, $^{190}$Pt. Indeed, the graphs corresponding to $^{182-186}$Pt are similar to that of $^{180}$Pt and those of $^{192-196}$Pt resemble that of  $^{190}$Pt. We may ask ourself why to make such plots once we know that the power expansion in $\gamma$ was performed around $\gamma=0^0$ and $\gamma=30^0.$ We notice that the density maxima are met not in the same point where the potential is minimum. The reason is that the density accounts also for the kinetic energy and moreover includes a factor defining the measure of the integration in the $\beta$ and $\gamma$ coordinates. These figures reflect the structure of the wave functions. Indeed, since the $\gamma$ dependent function depends on $\cos3\gamma$ and the spheroidal functions are symmetric with respect to the space reflection transformation, the graphs exhibit the symmetry $\gamma \to \pi/3-\gamma$. Concerning SMA the mentioned symmetry is caused by the fact the potential in $\gamma$ is function of $cos^23\gamma$. Also, the node of the $\beta$ function causes a doublet maxima with the same $\gamma$. For $^{188}$Pt we notice equal density curves which surround two maxima of identical beta. This situation is specific to the shape coexistence.  It is worth mentioning that such transition is showing up despite the fact that for all isotopes
$^{180-188}$Pt we used a power expansion in $\gamma$ around $0^0$. That means that the transition is caused not only by the potential shape but also by the structure coefficients involved in the associated differential equations. Actually, we calculated the spectroscopic properties of Pt isotopes with $A\ge 190$ also with a power expansion in $\gamma$ around $\gamma=0$. However, the results of SMA are characterized by a smaller r.m.s values for the deviations of the predictions from the experimental data.  It is interesting to note that although we changed the description when we passed from $^{188}$Pt to $^{190}$Pt the probability density undergoes a smooth transition. The maxima surrounded by equidensity curves merge in one maximum at $\gamma=30^0$ for ground and $\beta$ band states, while for $\gamma$ band states the doublets are well separated. How this picture is modified when additional degrees of freedom like octupole \cite{Ceau,Rad73} or single particle \cite{LimPL,LimZP} will be analyzed elsewhere.

Note that for $^{190}$Pt the considered excited state in the $\beta$ band is $8^+$ and not $10^+$ as happens for other isotopes. The reason is that, as seen from Table V, the highest spin state for which 
 energies in nuclei with $A\geq 190$, calculated with SMA, is $8^+$. 
\begin{figure}
\begin{center}
\includegraphics[height=21cm,width=14cm]{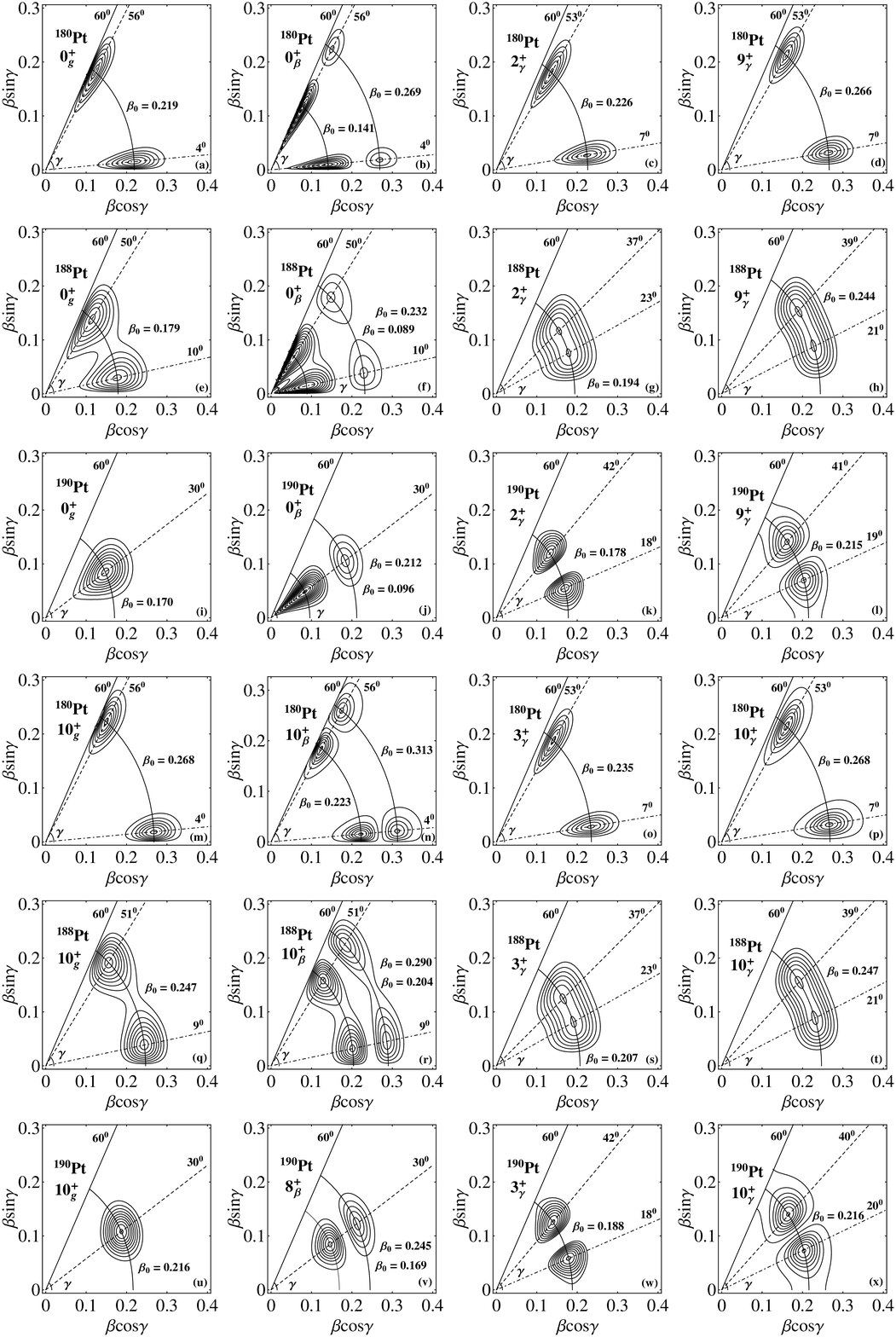}
\end{center}
\caption{(Color online) Probability densities for the states $0_{g}^{+}$, $10_{g}^{+}$, $0_{\beta}$, $10_{\beta}$, $2_{\gamma}^{+}$, $3_{\gamma}^{+}$, $9_{\gamma}^{+}$ and $10_{\gamma}^{+}$  of  $^{180,188}$Pt and $^{190}$Pt, calculated with SSA and SMA, respectively. The steps used in the contour plots are 30, 10, 20 for $^{180}$Pt, $^{188}$Pt and $^{190}$Pt, respectively. Exceptions are $0^+_{\beta}$ for $^{188}$Pt and $8^+_{\beta}$ for $^{190}$Pt, where  the steps are 12 and 25, respectively. }
\label{fig-1}
\end{figure}
\clearpage

\renewcommand{\theequation}{4.\arabic{equation}}
\setcounter{equation}{0}
\section{Conclusions}

In the previous Section we described some even-even isotopes of Pt by four solvable models emerging from the generalized Bohr Mottelson Hamiltonian. Indeed, for 
the isotopes with $180\le A\le 188$ the approaches are those abbreviated by SSA and ISWSA, respectively, while for the rest of nuclei, $190\le A\le 196$, the SMA and ISWMA are alternatively used. 
It is worth mentioning that the approach called ISWMA was used for the first time in the present paper.
Since the first set exhibits some features of the 
X(5) "symmetry" we compared the results of our calculations with those obtained with the X(5) formalism, if they are available. As for the other isotopes the results were compared with the 
Z(5) results. One concludes that our results are slightly better than those obtained with X(5) and Z(5) methods regarding both the excitation energies and reduced E2 transition probabilities.

The wave function structure is nicely reflected in the contour  plots for the probability density. It is suggested that due to the Hamiltonian symmetries the wave functions might be suitable for accounting for shape evolution as well as for  possible shape coexistence. 

{\bf Acknowledgment.} This work was supported by the Romanian Ministry for Education Research Youth and Sport through the CNCSIS project ID-2/5.10.2011.

\end{document}